\documentclass{COMNET}

\usepackage{cite} 
\usepackage{graphicx} 
\usepackage{caption}
\usepackage{url}
\usepackage{algorithm} 
\usepackage{algpseudocode}
\usepackage[dvipsnames]{xcolor}

\definecolor{yacoubcolor}{rgb}{0.98, 0.27, 0.62}
\begin{document}

\title{An Adaptive Bounded-Confidence Model of Opinion Dynamics on Networks}

\shorttitle{An Adaptive Bounded-Confidence Model of Opinion Dynamics on Networks} 
\shortauthorlist{U. Kan, M. Feng, and M. A. Porter} 

\author{
\name{Unchitta Kan}
\address{Department of Computational and Data Sciences, George Mason University} 
\name{Michelle Feng}
\address{Department of Computing + Mathematical Sciences, California Institute of Technology}
\and
\name{Mason A. Porter}
\address{Department of Mathematics, University of California, Los Angeles and Santa Fe Institute}
}

\maketitle

\begin{abstract}
{Individuals who interact with each other in social networks often exchange ideas and influence each other's opinions. A popular approach to study the spread of opinions on networks is by examining bounded-confidence models (BCMs), in which the nodes of a network have continuous-valued states that encode their opinions and are receptive to other nodes' opinions when they lie within some confidence bound of their own opinion. In this paper, we extend the Deffuant--Weisbuch (DW) model, which is a well-known BCM, by examining the spread of opinions that coevolve with network structure. We propose an adaptive variant of the DW model in which the nodes of a network can (1) alter their opinions when they interact with neighboring nodes and (2) break connections with neighbors based on an opinion tolerance threshold and then form new connections following the principle of homophily. This opinion tolerance threshold determines whether or not the opinions of adjacent nodes are sufficiently different to be viewed as `discordant'. Using numerical simulations, we find that our adaptive DW model requires a larger confidence bound than a baseline DW model for the nodes of a network to achieve a consensus opinion. In one region of parameter space, we observe `pseudo-consensus' steady states, in which there exist multiple subclusters of an opinion cluster with opinions that differ from each other by a small amount. In our simulations, we also examine the importance of early-time dynamics and nodes with initially moderate opinions for achieving consensus. Additionally, we explore the effects of coevolution on the convergence time of our BCM.
} 
{opinion dynamics, bounded-confidence models, coevolving networks, homophily}
\end{abstract}



\section{Introduction}

When individuals in a social network interact with each other, they often discuss and exchange ideas, and they can thereby influence each other's opinions. In social networks, similar individuals are more likely than dissimilar individuals to engage with each other \cite{mcpherson2001}. Such homophilic behavior can lead to homophilic communities and the formation of `echo chambers' on social and political issues \cite{kossinets2009,flaxman2016}. Homophily is thus a key lens to use when studying the formation and spread of opinions on social networks \cite{sasahara2020social,ortiz2021}. See \cite{khanam2020homophily} for a review of homophily in social network analysis.

Researchers have developed many models of opinion dynamics \cite{loreto2009,noor2020}, and network structure can significantly influence such dynamics \cite{porter2016,bullo2020}. In a model of opinion dynamics, the opinions of the nodes (which represent agents, such as individual humans or other entities) of a network can take either discrete values (such as in a classical voter model \cite{holley1975}, in which there are two possible node opinions) or continuous values. Well-known examples of the latter are \textit{bounded-confidence models} (BCMs) \cite{noor2020}, in which opinions take values either in an interval or in a higher-dimensional space. Allowing the opinions of agents to be points in an interval is useful for modeling opinions on a liberal--conservative political spectrum or on a single issue, and similar interpretations are possible when one considers two or more opinion dimensions \cite{brooks2020}. 

In a BCM, when two agents interact, they update their opinions by compromising by some amount if the difference between their opinions is below a specified threshold (i.e. if it is below a `confidence bound'). Otherwise, following standard practice, we suppose that the two agents do not adjust their opinions when they interact with each other.\footnote{Another possibility is for agents to `dig in their heels' and perhaps adjust their opinions so that they are farther apart from each other. For example, see one of the models in \cite{delvicario2017}.} One way to interpret an agent's confidence bound is as its willingness to engage with agents with different ideologies. (For example, perhaps it encodes their open-mindedness.) The two most famous BCMs are the Deffuant--Weisbuch (DW) model \cite{deffuant2000} and the Hegselmann--Krause (HK) model \cite{heg2000}. In the DW model, which we generalize in the present paper, one considers asynchronous updates of agent opinions and randomly chooses a single node pair (i.e. a dyad of agents) to interact in each time step. One then applies the opinion-update rule that we described above. In the HK model, agent opinions update synchronously.
 
Since the pioneering works of Deffuant et al. \cite{deffuant2000} and Hegselmann and Krause \cite{heg2000}, there have been many studies of BCMs (see, e.g., \cite{weisbuch2002, lorenz2007, lorenz2009, meng2018, sasahara2020social}), typically in the form of numerical investigations. Researchers have extended BCMs in a variety of ways, such as by incorporating content sharing and media nodes \cite{brooks2020}, updating opinions based on the median opinion (instead of the mean opinion) of interacting agents \cite{bullo-median}, and examining polyadic interactions of agents instead of only dyadic ones \cite{hickok2022}.

In the present paper, we generalize the DW model by formulating and studying an adaptive network model in which bounded-confidence opinion dynamics coevolve with network structure. Adaptive models of opinion dynamics give a convenient framework to examine the effects of opinion tolerance in idealized settings. There has been much research on opinion models in adaptive networks \cite{gross2007adaptive,sayama2013modeling}, especially in the form of adaptive voter models and their extensions (see, for example, 
\cite{holme2006,durrett2012graph,malik2016,chu2021}). Two recent studies examined adaptive opinion models with homophilic rewiring~\cite{li2021PRE,li2022}. There has also been research on network rewiring in the context of opinion tolerance, such as in an Axelrod model of cultural dissemination~\cite{gracia2011}.

Adaptive BCMs have been studied in a variety of contexts~\cite{kozma2008, kozma2008a, parravano2016, delvicario2017,brede2019, sasahara2020social,milli2022}. Kozma and Barrat \cite{kozma2008, kozma2008a} examined DW dynamics on adaptive Erd\H{o}s--R\'{e}nyi (ER) networks. They modeled the coevolving network and opinion dynamics using a biased coin flip. In their model, one chooses a dyad uniformly at random. With probability $p$, if the difference in the opinions of the two agents in the dyad is larger than some confidence bound, the edge between those two agents detaches from one of them and rewires to an agent that one chooses uniformly at random. With probability $1 - p$, the agent opinions evolve according to the DW opinion-update rule. More recently, Parravano et al. \cite{parravano2016} studied an adaptive BCM with signed edges (so that agents can either be friends or be enemies), with sign changes that depend on the distances between agent opinions. Del Vicaro et al. \cite{delvicario2017} formulated two adaptive BCMs and used them to examine the coexistence of polarized opinions at steady state. Brede \cite{brede2019} used an adaptive BCM to study the active participation of agents (who deliberately seek out other agents to try to change the opinions of those agents) in consensus formation. Sasahara et al. \cite{sasahara2020social} refashioned a BCM to model opinion dynamics in the context of social-media platforms in which users can encounter messages, repost messages, and unfollow users. Very recently, Pansanella et al.~\cite{milli2022} used a bounded-confidence mechanism and network rewiring to exam algorithmic bias in networks.

In our adaptive BCM, we incorporate an opinion tolerance threshold and a rewiring mechanism that follows the principle of homophily. Our model thus incorporates two dynamical processes and has both a tolerance threshold and a confidence bound. In each time step, we break the edges between probabilistically selected neighboring agents whose opinions are too far apart (because they exceed the opinion tolerance threshold and are thus `discordant') and we update the opinions of the agents using a standard bounded-confidence mechanism. When an edge breaks, one of these agents connects to a new agent with a probability that depends on the similarity of their opinions (i.e. it is based on homophily). In our numerical simulations, which employ ER networks, we observe that our adaptive DW model requires a larger confidence bound than a baseline DW model for the agents of a network to achieve a consensus. In one region of parameter space, we observe the emergence of a `pseudo-consensus' state, in which we observe multiple subclusters within an `opinion cluster' (i.e. a set of agents with very similar or consensus opinions) that have opinions that differ from each other by a small amount. When the agents have a low tolerance for neighbors with different opinions than theirs, our adaptive DW model behaves differently from the standard DW model. When the confidence bound is small, our model has faster cluster formation (and hence a faster convergence time) than our baseline DW model. When the confidence bound is large, our model has long convergence times, which is the opposite of what occurs in the baseline DW model. 

Our study has some similarities to \cite{kozma2008, kozma2008a, delvicario2017}, but a crucial difference is that our rewiring rule is based on the sociological principle of homophily \cite{khanam2020homophily}. The connections that people form on social media \cite{aiello2012} (and elsewhere) are influenced heavily by homophily, and it is important to incorporate such ideas into models of coevolving networks. In our BCM, agents both compromise their opinions with like-minded agents (according to a bounded-confidence mechanism) and dissolve old connections and form new connections based on the similarities of their opinions.\footnote{It is also interesting to consider attributes (such as similar demographic characteristics or hobbies) other than opinions.} New edges form between two specified agents with a larger probability when their opinions are more similar, instead of agents rewiring to agents uniformly at random. This resembles studies in adaptive voter models that have compared the effects of rewiring to agents with exactly the same opinion to rewiring to agents with any opinion \cite{durrett2012graph}. Another distinctive feature of our adaptive BCM is that the confidence bound and opinion tolerance threshold are distinct parameters. The opinion tolerance threshold yields a notion of discordant edges.\footnote{The concept of discordant edges has led to important insights in the study of adaptive voter models \cite{durrett2012graph}.} 

Our paper proceeds as follows. In Section \ref{background}, we give background information about the DW model. In Section \ref{sec3}, we present our adaptive DW model and discuss pertinent details of our implementation of it. In Section \ref{sec:results}, we discuss the results of our numerical simulations of our model. We conclude in Section \ref{sec5}. Our code is available at \url{https://gitlab.com/unchitta/coevolving-bc}.


\section{The Deffuant--Weisbuch Model} \label{background}

We briefly review the Deffuant--Weisbuch (DW) model of bounded-confidence opinion dynamics \cite{deffuant2000}. Consider a network of agents in which each agent $i$ holds an opinion $x_i$ that changes with time. The vector of opinions of a set of $N$ agents is the `opinion profile' $X = (x_1, \ldots, x_N)$. In the original DW model, the agents mix completely, so the opinion dynamics occur on a complete graph. 

In a given time step (which has a duration $\Delta t$), we choose two agents, $i$ and $j$, uniformly at random to interact with each other. If their opinions, $x_i(t)$ and $x_j(t)$, at time $t$ satisfy $|x_i(t) - x_j(t)| < C$ for some confidence bound $C$ (which one can interpret as the open-mindedness of the agents), they compromise their opinions through the opinion-update rule
 \begin{align}\label{update}
    x_i(t+ \Delta t) &= x_i(t) + \alpha (x_j(t) - x_i(t)) \,,  \notag \\
    x_j(t+ \Delta t) &= x_j(t) + \alpha (x_i(t) - x_j(t)) \,,
\end{align}
where $\alpha \in (0,0.5]$ is a constant that is often called a `convergence parameter' because it influences the speed of convergence. The parameter $\alpha$ governs how much agents compromise when they update their opinions.

As is typical in studies of BCMs \cite{meng2018}, we assume that all agents have the same confidence bound. In this scenario, when the dynamics reach a steady state, researchers have observed the formation of consensus and fragmented clusters of opinions, depending on the value of the confidence bound. Situations with exactly 2 opinion clusters are `polarized' and situations with 3 or more opinion clusters are `fragmented'.  When agents are receptive to opinions that deviate more from theirs (i.e. when their confidence bound is larger), they are willing to compromise with more of the other agents. As one increases the confidence bound, there is a phase transition from polarized/fragmented states to consensus states \cite{lorenz2007}. In the adaptive DW model in \cite{kozma2008, kozma2008a}, the critical point of this phase transition occurs at a larger confidence bound than in the standard DW model. In the polarized and fragmented regimes, which occur when the confidence bound is small, there are peaks (which indicate the presence of multiple opinion clusters) in the steady-state distribution of opinions. The number of peaks and the distance between the consensus opinions of the clusters depend on the confidence bound \cite{deffuant2000}. The number of agents in a system affects the time to converge to a steady state.


\section{Our Adaptive DW Model} \label{sec3}

Consider an undirected network (i.e. a graph) $G=(V,E)$, where $V$ is the set of nodes in the network and $E \subseteq V \times V$ is the set of edges. The nodes represent agents, and the edges represent social or communication ties between agents. The network coevolves with the opinions of the nodes, so the set of edges can change with time; it is thus helpful to write $E = E(t)$. The node set $V$ is constant in time, and $N = |V|$ is the number of nodes in the network. Let $C \in [0,1]$ be the confidence bound, and let $x_i(t) \in [0,1]$ denote node $i$'s opinion about some issue. The set of \textit{discordant edges} is $E_d^{\beta}(t) = \{ (i,j) \in E(t) : \left| x_i(t) - x_j(t) \right| > \beta \}$. That is, $E_d^{\beta}(t)$ is the set of dyads in the graph $G = G(t)$ whose constituent nodes have opinions that are farther apart than the opinion tolerance threshold $\beta \in [0,1]$ at time $t$. We interpret the threshold $\beta$ as a homophily parameter. A larger value of $\beta$ entails a more stringent requirement for an edge to be discordant. In many situations, it seems intuitive to require that $\beta \geq C$ (which implies that an edge between two nodes can be discordant only when the difference between the nodes' opinions is at least as large as the confidence bound), but we do not include this requirement in our model. 

Time is discrete, and two processes occur in series in each time step. In the first process, nodes rewire; in the second process, nodes update their opinions. In the rewiring process, we choose up to $M$ discordant edges to rewire according to a homophilic rewiring rule that typically changes the network structure. We describe this mechanism in Section \ref{sec3.1}. We then choose $K$ dyads; the nodes in each dyad adjust their opinions according to the update rule \eqref{update} of the DW model \cite{deffuant2000} if the difference in their opinions is within a confidence bound. We give more details in Section \ref{sec3.2}.

In a real-world context, the first process may correspond to an individual `unfriending' one of their connections in a social network (e.g. on a social-media platform) and then establishing a connection with someone else. Such unfriending occurs when an individual cannot tolerate the difference between their opinion and that of the individual that they are unfriending. The individual then chooses to befriend someone else. Based on the principle of homophily, the individual is more likely to befriend somebody whose opinion is similar to theirs than somebody whose opinion differs greatly from theirs. We illustrate this process in Figure \ref{fig:rewiring_schematic}. After this rewiring process, we choose dyads for opinion updating; the agents in these independently-chosen dyads update their opinions according to \eqref{update} if the difference between their opinions lies within the confidence bound.\footnote{In our BCM, interacting agents compromise their opinions when the difference between their opinions is exactly equal to the confidence bound; the standard DW model requires this difference to be strictly less than the confidence bound.}

We simulate our model, with both processes occurring in each time step, until there are no further noticeable changes of the opinions of the agents of a network. In our computational experiments, we stop a simulation if the sum of the changes of the agent opinions is less than $\text{tol} = 10^{-5}$ for each of $100$ consecutive time steps. This is our numerical tolerance for convergence. Because it is possible that a simulation may fail to satisfy this termination criterion, we set a `bail-out time' and stop a simulation after $10^6$ time steps if it has not already stopped.

We give pseudocode for our adaptive DW model in Algorithm \ref{pseudocode}. Our code is available at \url{https://gitlab.com/unchitta/coevolving-bc}.


\subsection{Homophilic Rewiring} \label{sec3.1}

\begin{figure}[t!] 
\centering
\includegraphics[width=0.9\textwidth]{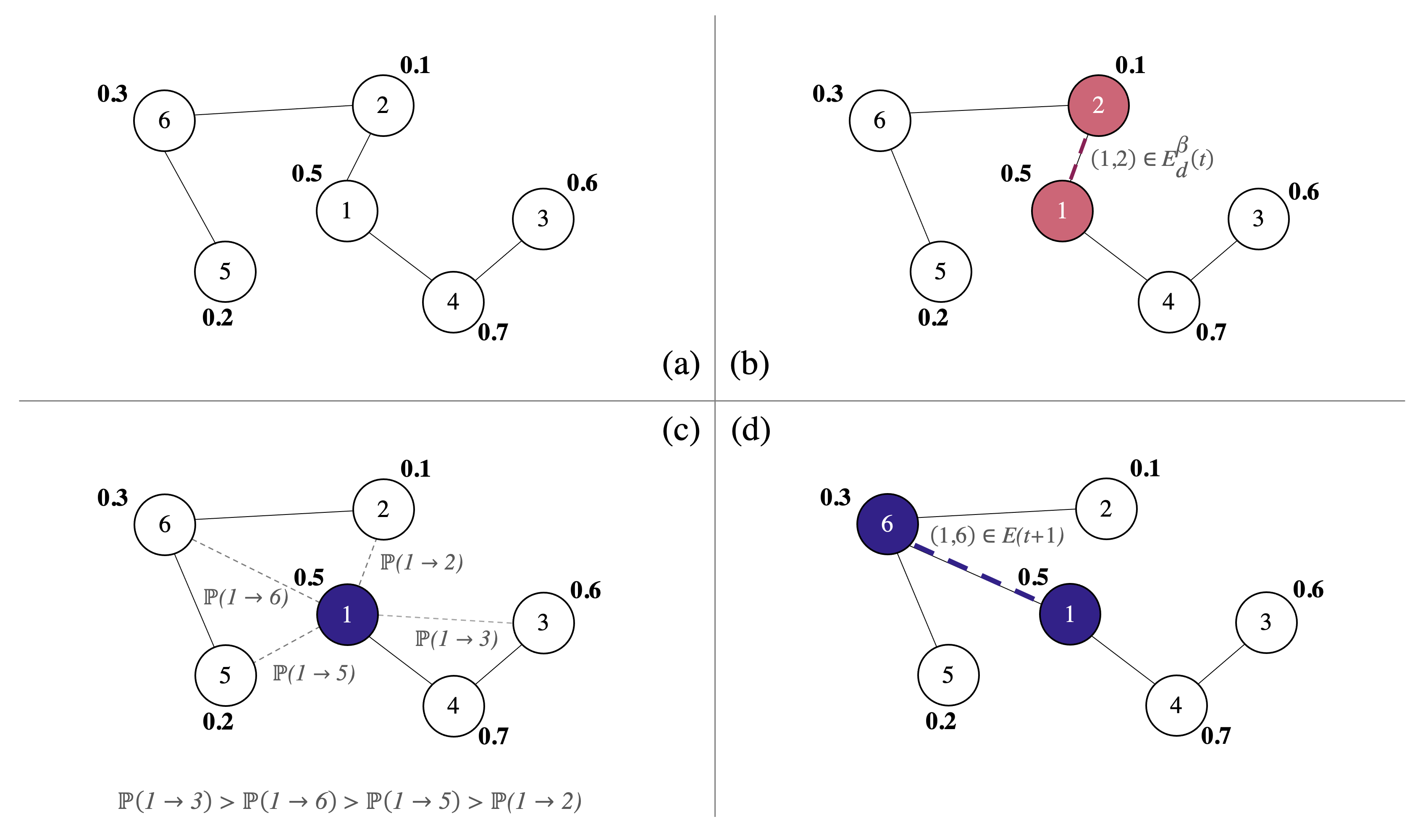}
\caption{A schematic illustration of the rewiring mechanism in our adaptive DW model. In this example, the confidence bound is $C = 0.3$ and the opinion tolerance threshold is $\beta = 0.2$. In (a), we show the current network structure and node opinions. For example, node $1$ has opinion $0.5$. In the depicted time step, we choose the edge $(1,2)$ uniformly at random from the set of discordant edges and remove this edge (see (b)). As we illustrate in (c), we then choose node $1$ to rewire to a new node. 
(We select which of nodes $1$ or $2$ rewires with equal probability.) The rewiring probabilities (with $\mathbb{P}(i \rightarrow a)$ denoting the probability that node $i$ rewires to form an edge to node $a$) in this time step depend on the opinions of the nodes. In this example, we form the new edge $(1,6)$ (see (d)).
} \label{fig:rewiring_schematic}
\end{figure}

\begin{algorithm}[t!]
	\caption{Pseudocode for our Adaptive DW Model}
	\hspace{-.2 in} \textbf{parameters: } $N$, $p$, $M$, $K$, $\alpha$, $\beta$, $C$
	\begin{algorithmic}[1]
       \State $t \leftarrow 0$; $G \leftarrow G(N,p)$
	   \For {$i \in G\text{.nodes()}$ }
	        \State $x_i(0) \leftarrow \text{Unif}[0,1]$
	   \EndFor 
		\While {(time $t<$ bail-out time) and (sum of the magnitudes of the opinion changes $<$ tol for fewer than 100 consecutive steps) }
		\State $E_d^{\beta}(t) \leftarrow \emptyset$ \, [initialize set of discordant edges]
		\For {$(i,j)$ in the set of edges $E(t)$ }
		    \If {$|x_i(t) - x_j(t)| < \beta$ } $E_d^{\beta}(t) \leftarrow E_d^{\beta}(t) \cup \{(i,j)\}$ \EndIf
		\EndFor
		\If {$|E_d^{\beta}(t)| > M$ }
		    select $M$ edges uniformly at random from $E_d^{\beta}(t)$
		\Else { select all edges from $E_d^{\beta}(t)$ } 
		\EndIf
		\For {each discordant edge $(i,j)$ }
			\State dissolve and remove the edge from the edge set $E(t)$
			\State select node $i$ or $j$ with equal probability
			\State compute probabilities to rewire to other nodes using \eqref{eq3.1}
			\State randomly pick another node using the computed rewiring probabilities
			\State connect the node to the previously selected node with an edge; add the new edge to $E(t)$
		\EndFor
		\State select $K$ edges uniformly at random from $E(t)$
		\For {each selected edge $(i,j)$ }
		    \If {$|x_i(t) - x_j(t)| \leq C$ }
		        update the opinions of the nodes using \eqref{eq3.2}   \\ \hspace{2 in} [see the main text for further discussion]
		    \EndIf
        \EndFor
	   \State compute the sum of the magnitudes of the opinion changes
	   \State $t \leftarrow t + \Delta t$
	   \EndWhile
	\end{algorithmic} 
	\label{pseudocode}
\end{algorithm}

In our BCM's rewiring process, we select $M$ edges uniformly at random from the set $E_d^{\beta}(t)$ of discordant edges. If $|E_d^{\beta}(t)| \leq M$, then we select all of the discordant edges. We remove each edge $(i,j)$ in this set, and we select node $i$ or $j$ with equal probability to form an edge to a new node $a$. We choose the node $a$ with a probability that depends on the similarity between its opinion and the selected node's opinion. To work with mathematically well-defined similarities, we use a metric $d(x,y)$ on the space of node opinions to determine the similarities. One can also determine similarities using higher-dimensional opinion spaces or in a way that incorporates node attributes in addition to their current opinions (see, e.g., \cite{brooks2020}).

Suppose that we choose node $i$ from the dyad $(i,j)$. The probability that node $i$ rewires to a node $a$ is 
\begin{equation}\label{eq3.1}
    \mathbb{P}(i\rightarrow a) = \frac{1}{D}(1 - d(x_i,x_a))\,,
\end{equation}
where $d(x,y)$ is a metric (e.g. the $L_1$ norm or the $L_2$ norm) and we will shortly specify the normalization constant $D$. We use the $L_2$ norm, so $d = \| \cdot \|_2$. We set $\mathbb{P}(i\rightarrow i)$ to $0$ to prevent self-edges, and we set the probability that $i$ rewires to one of its existing neighbors to $0$ to avoid multi-edges. We include the constant $D$ in Eq.~\eqref{eq3.1} because we need a normalization to ensure that $\mathbb{P}(i\rightarrow a)$ is a probability. We calculate $D$ by solving $\sum_{z} \mathbb{P}(i\rightarrow z) = \frac{1}{D} \sum_{z} \left[ 1-d(x_i,x_z) \right]= 1$.

After choosing a node $a$ randomly according to the probabilities in \eqref{eq3.1} (while avoiding self-edges and multi-edges), we add the new edge $(i,a)$ to $E(t+1)$. If the new edge is discordant, we also add it to $E_d^{\beta}(t)$. We allow $i$ to rewire to $j$, even though the former just unfriended the latter. Sometimes, life just works that way (and it is convenient for our computations).

In Figure \ref{fig:rewiring_schematic}, we illustrate the rewiring process with example values of the confidence bound $C$ and the opinion tolerance threshold $\beta$.


\subsection{Opinion Updates} \label{sec3.2}

After the rewiring step, we select $K$ dyads uniformly at random (without replacement), and we adjust the opinions of the nodes in each dyad using the DW opinion-update rule \eqref{update}. Specifically, at time $t$, two interacting nodes $i$ and $j$ adjust their opinions if the difference between their opinions is less than or equal to the confidence bound (i.e. if $|x_i(t) - x_j(t)| \leq C$). In contrast to the traditional DW model, interacting nodes compromise their opinions when the difference between their opinions is exactly equal to the confidence bound. Nodes $i$ and $j$ change their opinions according to the update rule
\begin{align} \label{eq3.2}
    x_i(t+1) &= x_i(t) + \alpha (x_j(t) - x_i(t))\,, \notag \\
    x_j(t+1) &= x_j(t) + \alpha (x_i(t) - x_j(t))\,,
\end{align}
where $\alpha$ is the convergence parameter. Equation \eqref{eq3.2} is the same as equation \eqref{update}, except that we now specify that the time step has duration $\Delta t =1$. (We use $\Delta t = 1$ in all of our numerical computations.) If the opinions of nodes $i$ and $j$ are not within the confidence bound $C$, then they do not change their opinions. The parameters $M$ (i.e. the number of discordant edges that we rewire in one time step) and $K$ (i.e. the number of dyads that we consider when updating opinions) also affect the rate of convergence to a steady state. For example, a larger value of $K$ signifies that there are more encounters between agents in each time step, so more agents can compromise their opinions in a single time step. 

It is possible for the same node $h$ to be in more than $1$ of the $K$ dyads in a time step. We consider the $K$ dyads in an order that we select uniformly at random. For a given node, we apply only the final opinion update \eqref{eq3.2} to the opinion that it holds at the beginning of this time step. For example, suppose that we first select a dyad with nodes 1 and 2 and that we then select a dyad with nodes 2 and 3. Additionally, suppose that the pairwise differences between the node opinions all lie within the confidence bound. At the conclusion of the time step, node 1 has adjusted its opinion based on its interaction with node 2, but node 2 has adjusted its opinion only its interaction with node 3. Such an asymmetry in opinion updates cannot occur in the standard DW model. Additionally, because of our convention and the associated asymmetry in opinion updates, the sum of the opinions of the agents is not a conserved quantity in our adaptive DW model.

We run our simulations --- with first the rewiring process and then the opinion-update process in each time step --- until the system converges to a steady state (within the numerical tolerance level that we discussed previously) or until we reach the bail-out time. The steady state can include one or more opinion clusters. The quantity $M/K$ determines the relative time scales of the rewiring dynamics and the opinion dynamics. When $M/K \ll 1$, the opinions of the nodes change much faster than the network structure; by contrast, when $M/K \gg 1$, network structure changes much faster than the node opinions.


\section{Numerical Simulations}  \label{sec:results}

\begin{figure}[t!] 
\centering
\includegraphics[width=\textwidth]{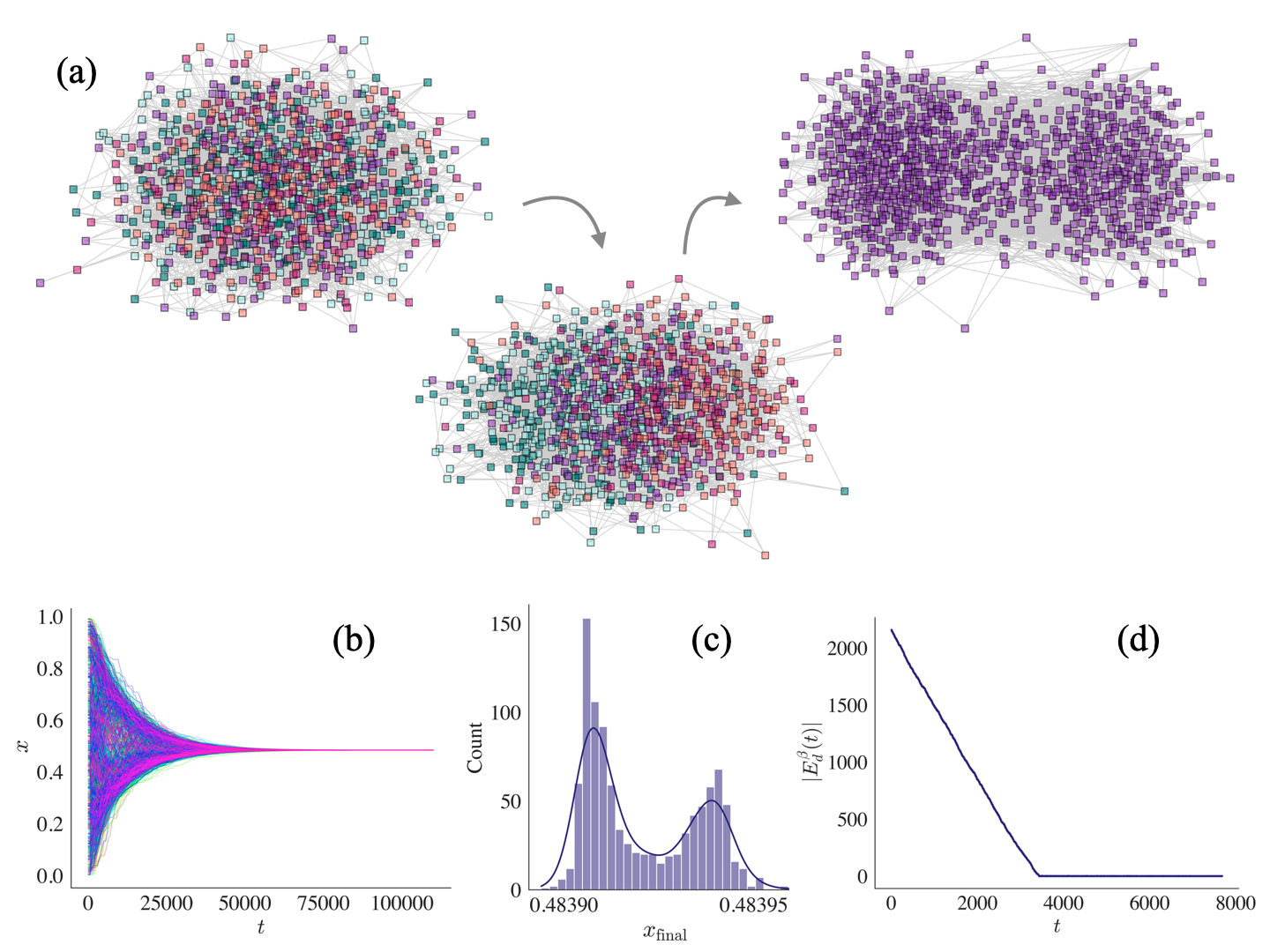}
\caption{Overview of one simulation of our adaptive DW model with a confidence bound of $C = 0.24$ and an opinion tolerance threshold of $\beta = 0.32$. We show (a) three snapshots of the network structure as it changes with time, (b) the time evolution of the opinions of the nodes in the network, (c) the distribution of the node opinions at steady state, and (d) the number of discordant edges as a function of time. The colors in (a) and (b) indicate opinion values.} 
\label{fig:overview}
\end{figure}

Because we need to consider both changes in opinions and changes in network structure, it is natural to ask how these changes affect each other. We explore the interaction between these two aspects of our model's dynamics by examining both network topology and node opinions as a function of time. We present numerical results of simulations of our adaptive DW model on synthetic networks that we generate using the $G(N,p)$ ER random-graph model \cite{newman2018networks}. It is useful to recall the parameters of our adaptive DW model: the number $N$ of nodes of a network, the number $M$ of edges that we rewire in one time step, the number $K$ of dyads that we consider when updating opinions, the convergence parameter $\alpha$ of the DW opinion-update rule \eqref{eq3.2}, the confidence bound $C$, and the opinion tolerance threshold $\beta$. For a network that we construct using the $G(N,p)$ model, the connection probability between nodes is $p = \langle k \rangle/N$, where $\langle k \rangle $ is the expected mean degree of the network. We set $p = 0.01$ (for which $\langle k \rangle = 10$), $M/K = 1/5$, and $\alpha = 0.1$. We draw the initial opinion $x_i(0)$ of each node $i$ randomly from the uniform distribution $\mathrm{Unif}[0,1]$, so the initial opinion profile is $X(0) \in [0,1]^N$. In Figure \ref{fig:overview}, we show the results of a single simulation of our model with $C = 0.24$ and $\beta = 0.32$. 

In most of our numerical computations (see Sections \ref{sec:4.1}--\ref{sec:4.4}), we consider networks with $N = 1000$ nodes. For these simulations, we use $M = 1$ and $K = 5$. In these simulations, for each examined location in the $(\beta, C)$ parameter plane, we do 50 independent simulations\footnote{For $(\beta, C) = (0.42,0.22)$, there are only 49 simulations.} and report sample means of them. All of our results with $N = 1000$ use the same set of simulations (sometimes from particular locations in the parameter plane). To explore finite-size effects, we also briefly examine our model on networks with $N = 5000$ nodes (see Section \ref{sec:4.5}). In these simulations, it is still the case that $M/K = 1/5$, but we now use $M = 5$ and $K = 25$. We still set $p = 0.01$, so the expected mean degree of our networks is now $\langle k \rangle = 50$.


\subsection{Concepts, Definitions, and Other Specifications}\label{sec:4.1}

Before presenting our computational results, we outline several concepts that are helpful for understanding the behavior of our model. In the standard DW model, agents are in consensus when they have the same opinion. A state with 2 opinion clusters is polarized, and a state with at least 3 opinion clusters is fragmented. In our adaptive DW model, it is convenient to relax these notions a bit. For example, our notion of consensus includes situations in which nodes are almost (but not perfectly) in agreement. When we observe a polarized or fragmented steady state in a network, we often also observe that the network has multiple connected components.


\subsubsection{Transitions Between Polarization/Fragmentation and Consensus}

In the standard DW model, when the confidence bound $C$ is small and below some critical value $C'$, one obtains a steady state with multiple opinion clusters. Intuitively, when $C$ is small, the agents of a network are close-minded and only interact with like-minded agents, so opinion clusters emerge. These opinion clusters are in different areas of the space of opinions, so one can interpret them as echo chambers in a social network \cite{flaxman2016}. Typically, we observe more opinion clusters as we consider more close-minded agents. As agents become more open-minded (i.e. for larger values of $C$), more agents engage with each other, which leads to more compromises and less fragmentation into different opinion clusters. If agents are sufficiently open-minded (specifically, if $C > C'$), they can achieve a consensus, so our adaptive DW model appears to have a phase transition at the critical value $C'$. Because $C'$ can depend on $\beta$, we denote putative transition values in our model by $C'_{\beta}$


\subsubsection{Polarization and Fragmentation}

When consensus does not occur, it is useful to further characterize the opinion clusters. We use the term `major cluster' for any opinion cluster that includes at least $5$ \% of the nodes of a network (e.g. at least $50$ of $N = 1000$ nodes) and `minor cluster' for any opinion cluster with fewer than $5$ \% of the nodes. We use the term `polarization' for situations with exactly 2 major clusters at steady state and `fragmentation' for situations with 3 or more major clusters at steady state.


\subsubsection{Consensus and Pseudo-Consensus}

In a consensus regime, there is only 1 major opinion cluster, which either has a single consensus opinion or has opinions that are almost in perfect consensus. In particular, we observe situations in which (upon close inspection) a major cluster has subclusters with opinions that differ from each other by a small value $\epsilon$. In our model, this situation has an associated community structure. We refer to this type of consensus as a `pseudo-consensus'. In all examples that we checked manually, the pseudo-consensus opinion clusters have exactly two major subclusters.


\subsubsection{`Small', `Intermediate', and `Large' Values of $\beta$ and $C$}

When discussing the results of our simulations, we often describe the values of the opinion tolerance threshold $\beta$ and the confidence bound $C$ as `small', `intermediate', or `large'. When $\beta$ is small, agents are more aggressive at cutting ties with agents whose opinions differ from theirs and then befriending agents whose opinions are similar to theirs. When $C$ is small, agents are close-minded and are influenced only by agents whose opinions are similar to theirs. These two key parameters affect the time scales of the rewiring and the opinion updates.

As we will see in our numerical experiments, different combinations of $\beta$ and $C$ can lead to very different behaviors at steady state. In practice, we observe that agents cut ties frequently when $\beta \lessapprox 0.2$, so we refer to such values of $\beta$ as `small'. We observe that agents are close-minded when $C \lessapprox 0.2$, so we refer to such values of $C$ as `small'. Additionally, we view $\beta$ as `intermediate' when $0.2 \lessapprox \beta \lessapprox 0.4$ and $C$ as `intermediate' when $0.2 \lessapprox C \lessapprox 0.3$. Finally, we view $\beta$ as `large' when $\beta \gtrapprox 0.4$ and $C$ as `large' when $C \gtrapprox 0.3$.


\subsection{Baseline Case (i.e. $\beta = 1$)} \label{sec:4.2}

\begin{figure}[t!] 
\centering
\includegraphics[width=0.95\textwidth]{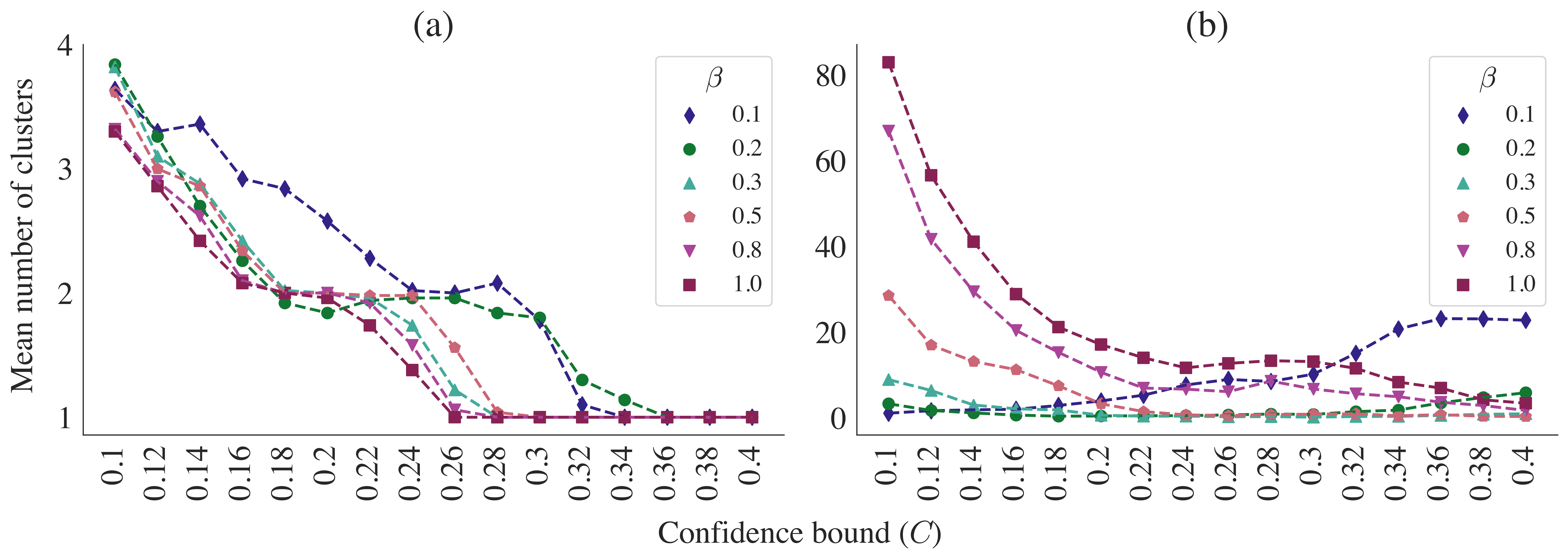}
\caption{The mean numbers of (a) major opinion clusters and (b) minor opinion clusters in our adaptive DW model at steady state as a function of the confidence bound $C$ for different values of the opinion tolerance threshold $\beta$. The results are means of $50$ simulations. We initialize each simulation with an independently generated ER graph (with $N = 1000$ nodes) and a distinct opinion profile that we draw from the uniform distribution $\mathrm{Unif}[0,1]$. We use the same set of simulations (sometimes from particular locations in the $(\beta, C)$ parameter plane) for all of our results with $N = 1000$ nodes. 
} 
\label{fig:num-clusters}
\end{figure}

\begin{figure}[t!] 
\centering
\includegraphics[width=0.58\textwidth]{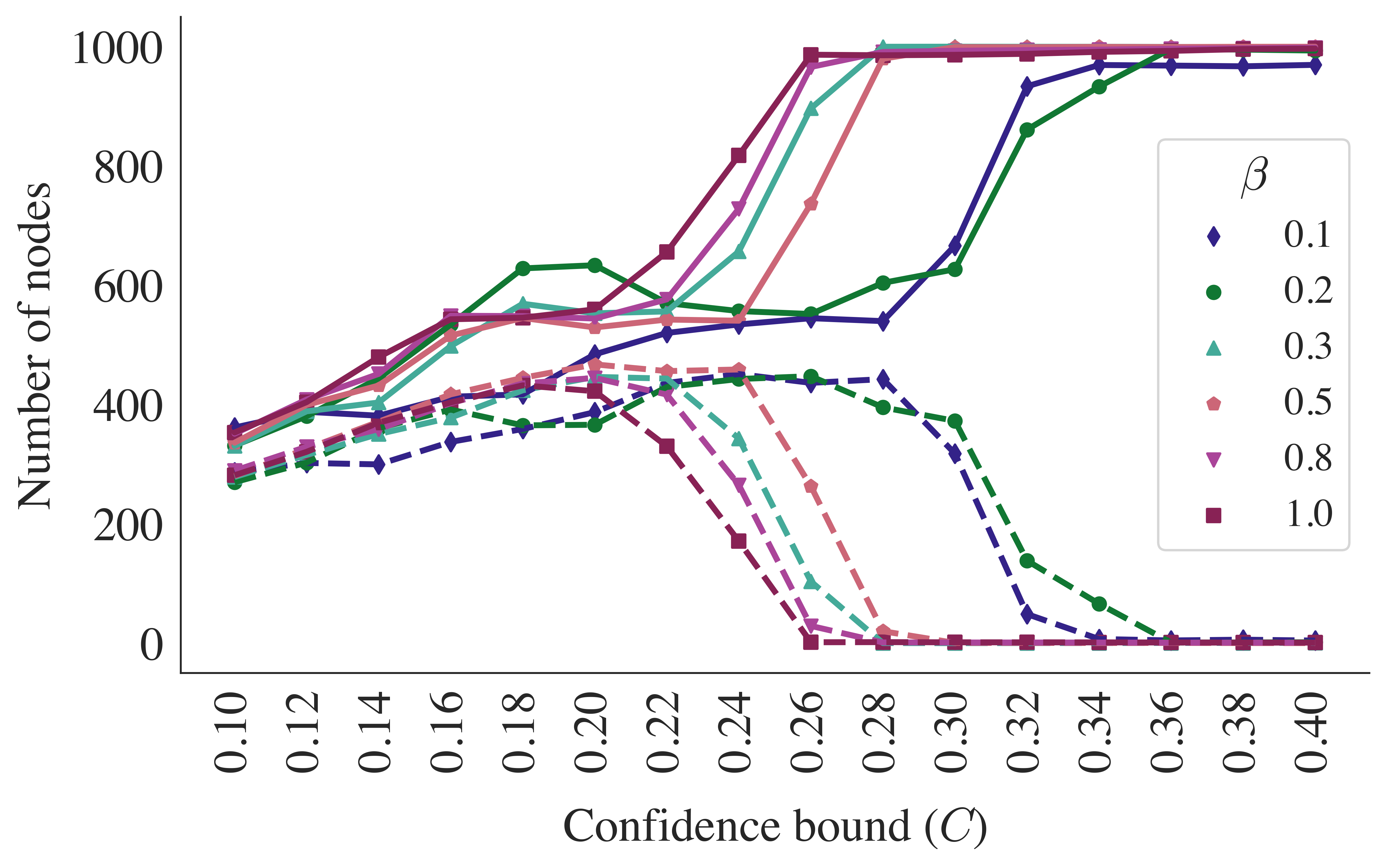}
\caption{The mean numbers of agents in the largest opinion cluster (solid curves) and the second-largest opinion cluster (dotted curves) in our adaptive DW model versus the confidence bound $C$ for different values of the opinion tolerance threshold $\beta$. The results are means of the same set of 50 simulations as in Figure \ref{fig:num-clusters}.} 
\label{fig:largest-clusters}
\end{figure}

\begin{figure}[t!] 
\centering
\includegraphics[width=0.9\textwidth]{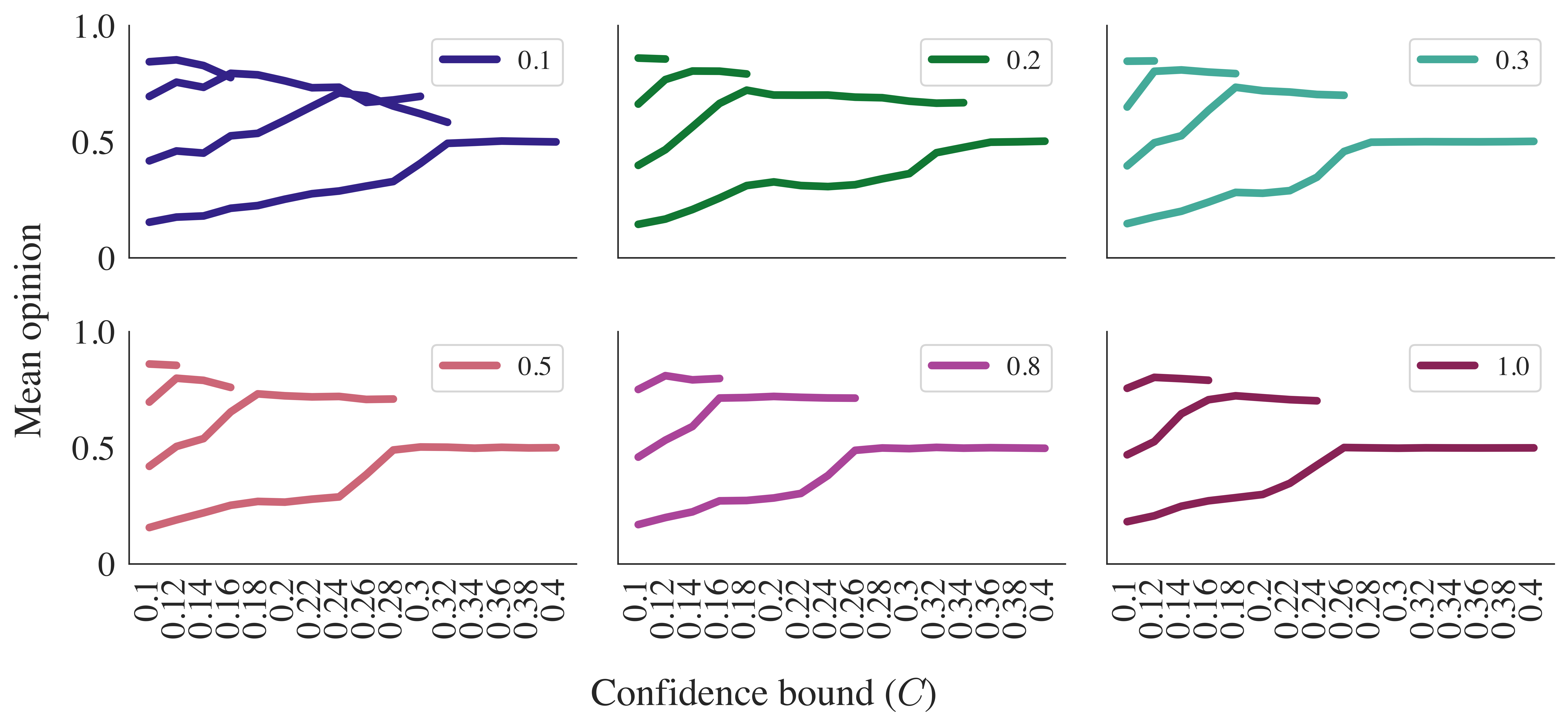}
\caption{The means of the opinions of the largest major opinion clusters in our adaptive DW model versus the confidence bound $C$. Each panel gives results for a different value of the opinion tolerance threshold $\beta$; we show up to four clusters for each value of $\beta$. (We show three major clusters in situations where there are only three such clusters.)
The results are means of the same set of 50 simulations as in Figure \ref{fig:num-clusters}.} 
\label{fig:avg-opinion}
\end{figure}

When $\beta=1$ in our adaptive DW model, there is no rewiring because the set of discordant edges is always empty, so our model reduces to a variant of the standard DW model. Therefore, we treat the results that we obtain with $\beta = 1$ as a baseline. Our baseline DW model behaves like the standard DW model, but it has faster convergence times because more of its nodes can change opinions in a single time step. 

The baseline (i.e. non-coevolving) DW model appears to have a phase transition between consensus and polarized/fragmented steady states at a critical value of the confidence bound. For the standard DW model on a complete network, numerical computations in previous works suggest that this critical value is $C' \approx 0.26$ \cite{deffuant2000, lorenz2007}. Kozma and Barrat~\cite{kozma2008} obtained $C' \approx 0.256$ in simulations of a DW model on time-independent ER networks. In our adaptive DW model, $C'$ depends on $\beta$, so we denote putative transition values by $C'_{\beta}$. In our baseline model (i.e. when $\beta = 1$), we observe a transition between consensus and polarization at $C'_{\beta = 1} \approx 0.26$. 

In Figure \ref{fig:num-clusters}(a), we show the mean numbers of major opinion clusters at steady state in simulations of our model on ER networks with $1000$ nodes. The opinion profile fragments into 2--4 major clusters when $C \in [0.1, 0.16)$, and it is polarized when $C \in [0.16,0.26)$. When $C \gtrapprox 0.26$, the system reaches a consensus, except for very small (i.e. minor) opinion clusters (some of which consist of isolated nodes). In Figure \ref{fig:largest-clusters}, we show the mean number of nodes in the largest opinion cluster at steady state; in the polarized regime, the two largest opinion clusters each have almost $500$ nodes. In Figure \ref{fig:avg-opinion}, we show the mean opinion of the largest major opinion clusters as a function of the confidence bound $C$.


\subsection{Homophilic Rewiring (i.e. $\beta < 1$)} \label{sec:4.3}


\subsubsection{Shifted Critical Value (i.e. Shifted Phase Transition)}

When $\beta < 1$ (i.e. when there is homophilic rewiring), and especially when $\beta$ is small, we often observe that the transition between polarization/fragmentation and consensus occurs at a larger value of $C$ than for the baseline DW model (i.e. when $\beta = 1$). That is, consensus is harder to achieve when agents are intolerant of opinions that differ much from theirs, as edges are more readily discordant and nodes thus rewire more often. In particular, although we observe a consensus steady state when $C \gtrapprox 0.26$ in the baseline DW model, our adaptive DW model results in a polarized or fragmented steady state up to $C'_{\beta} \approx 0.34$ when $\beta$ is small. Additionally, as we see in Figure \ref{fig:largest-clusters}, the transition value $C'_{\beta}$ depends on $\beta$ in a complicated way. For example, the simulations of our adaptive DW model with $\beta \approx 0.3$ behave rather differently than those with $\beta \approx 0.2$.


\subsubsection{Pseudo-Consensus: Competing Time Scales and the Role of Moderate Agents}

By allowing nodes to break connections based on an opinion tolerance threshold, we also observe other interesting phenomena. For example, agents can still sometimes achieve a consensus --- although it is sometimes technically in the form of a pseudo-consensus --- when $C < C'_{\beta}$ for certain values of $\beta$. Figure \ref{fig:overview} shows an example of a pseudo-consensus state. In this example, opinions seemingly converge to one value near the center of the opinion space (see Figure \ref{fig:overview}(b)). However, upon closer inspection (see Figure \ref{fig:overview}(c)), we observe subclusters of the steady-state opinion cluster that differ from each other by a small value $\epsilon \approx 0.0003$. Our visualization in Figure \ref{fig:overview}(a) also suggests that there is some community structure. 

It seems that a pseudo-consensus can arise from the presence of two competing processes (rewiring and opinion changes) with comparable rates. We consider the interactions of these processes by examining the relationship between the confidence bound $C$ and the opinion tolerance threshold $\beta$. Additionally, the value of $M/K$ affects the relative rates of rewiring and opinion changes in a network. When there is rewiring, our BCM tends to yield homophilic communities (including situations in which networks themselves fragment into multiple components), which one can interpret as echo chambers. The opinion updates for $C \gtrapprox 0.24$ tend to encourage consensus because agents are open-minded, so many opinion compromises can occur. If a network organizes into distinct communities faster than agents achieve a consensus, we expect to observe polarization or fragmentation. However, if the agents of a network reach a consensus faster than the network rewires into poorly connected or disconnected communities, we expect the system to reach consensus. When the processes have comparable rates, a pseudo-consensus may arise. In this case, we observe situations in which there is one major opinion cluster, but there appears to be community structure in the cluster (see Figure \ref{fig:overview}(a)). 

Consider the case in which homophilic rewiring is faster than opinion changes. Specifically, suppose that both $\beta$ and $C$ are small. When $C$ is small (e.g. $C < 0.2$), nodes tend to begin with fewer neighbors that can influence them than for larger values of $C$, so their opinions change much more slowly at the beginning of a simulation. Calculating the total number of `influential neighbors' (i.e. the number of neighbors whose opinion is within $C$ of a node) of each node versus time confirms this observation. If $\beta$ is also small (e.g. $\beta < 0.2$), nodes can quickly disconnect from neighbors whose opinions are too far away from theirs and attach to nodes whose opinions are closer to theirs. This leads to the formation of communities or even multiple connected components in a network. Once the nodes of a network organize into such homophily-based communities, they then compromise their opinions within these communities and quickly achieve an intra-community consensus. (The opinions of the nodes in a community are likely to be within one another's confidence bounds, so two nodes in the same community tend to compromise when they interact.) By this time, there are very few or even no remaining discordant edges, so these communities persist over time. However, when the opinions of a network's nodes converge very quickly (specifically, for large $C$), the nodes do not have many chances to disconnect from their discordant neighbors to form communities. Therefore, it is reasonable that a pseudo-consensus (i.e. a consensus with subclusters and community structure) emerges when the rates of opinion changes and rewiring are similar. In Figure \ref{fig:avg-convergence-time}, which we discuss in detail in Section \ref{sec:4.4}, we identify the region in the $(\beta, C)$ parameter plane in which we often observe pseudo-consensus. 

A small number of nodes in an opinion cluster have opinions that lie somewhere between those in the subclusters that we just identified. Such agents often have initially moderate opinions (e.g. $x_i \approx 0.5 \pm 0.05$). We hypothesize that (1) these initially moderate agents are crucial to allow the agents of a network to reach a pseudo-consensus instead of becoming polarized or fragmented and (2) initially moderate agents act like `bridges' that keep a network connected. In our model, when we observe a polarized or fragmented steady state in a network, we often also observe that the network has multiple connected components.


\subsubsection{Minor Opinion Clusters: Isolated and Extreme Agents}

In Figure \ref{fig:num-clusters}(b), we observed that the the number of minor opinion clusters is very different when there is rewiring (i.e. when $\beta < 1$) than when there is not (i.e. when $\beta = 1$). Notably, there are very few minor opinion clusters when $C$ and $\beta$ are both small. When $C$ is small, agents tend to be influenced only by a few other agents and thus are likely to form small groups or even become isolated. However, when there is rewiring, agents are able to replace neighbors with discordant opinions from theirs with like-minded agents that potentially can become part of the same major opinion cluster.

When the opinion tolerance threshold $\beta$ is very small (e.g. $\beta = 0.1$), the number of minor opinion clusters at steady state increases with the confidence bound $C$. These minor clusters tend to consist of isolated nodes or of 2--3 nodes. Most of the agents in these minor clusters have initially extreme opinions (i.e. close to $0$ or close to $1$). The fast formation of a moderate-opinion consensus and the very small value of $\beta$ causes agents to disconnect from agents with extreme opinions (i.e. `extreme agents') very early in simulations when they become more moderate. The agents that remain extreme are in minor clusters at steady state. 

The outcomes that we observe in our simulations in the presence of homophilic rewiring have real-world analogues. For example, when $\beta$ is small, it is reasonably common in our simulations for small minority groups of extreme agents to be sparsely connected to other parts of a network. These extreme agents cannot be swayed by a majority with more moderate opinions. In the real world, it seems that many people in a population also may dissociate from unpopular minorities. Interestingly, however, the number of such extreme minority groups increases when we increase the open-mindedness of the agents in our model.


\subsubsection{Homophily} 

Because we base the rewiring process in our adaptive DW model on the principle of homophily, it is sensible to compute a measure of homophily in our networks. For simplicity, we calculate the scalar assortativity coefficient \cite{newman2018networks}
\begin{equation} \label{eq:assortativity}
	r = \frac{\sum_{ij}\left(A_{ij} - \frac{k_{i}k_{j}}{2m}\right)x_{i}x_{j}}{\sum_{ij}\left(k_{i}\delta_{ij} - \frac{k_{i}k_{j}}{2m}\right)x_{i}x_{j}} \,,
\end{equation}
where $x_i$ is an ordered scalar attribute that is associated with node $i$, the scalar $k_i$ is node $i$'s degree, $m$ is the total number of edges of a network, and $\delta_{ij}$ is the Kronecker delta function. Equation \eqref{eq:assortativity} is a normalized, network-based generalization of the Pearson correlation efficient. Its value ranges between $-1$ and $1$, and it equals $0$ either when the attribute has only one value or when there is no linear correlation between the attribute values at adjacent nodes. When $r = 1$, the network is perfectly assortative; when $r = -1$, it is perfectly disassortative.

\begin{figure}[bt!]
\centering
\includegraphics[width=1.0\textwidth]{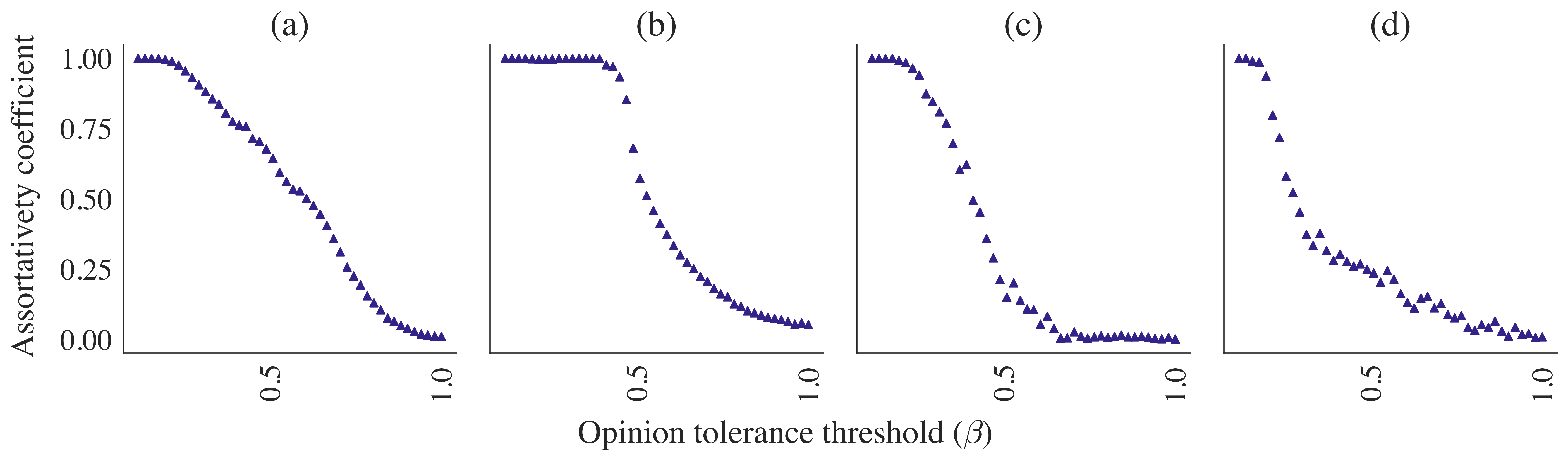}
\caption{The assortativity coefficient $r$ for the final opinions of the agents in our adaptive DW model as a function of the opinion tolerance threshold $\beta$ for confidence bounds of (a) $C = 0.1$, (b) $C = 0.2$, (c) $C = 0.3$, and (d) $C = 0.4$. The results are means of the same set of 50 simulations as in Figure \ref{fig:num-clusters}.
}
\label{fig:assortativity}
\end{figure}

In Figure \ref{fig:assortativity}, we show mean values of the assortativity coefficient $r$ of opinions in the final networks of our set of $50$ simulations. For each simulation, $x_i$ is the opinion of node $i$ at the end of the simulation. As expected, for any confidence bound $C$, the values of $r$ are large when the opinion tolerance threshold $\beta$ is small. Although $r \approx 1$ (i.e. almost perfectly assortative mixing) for all values of $C$ when $\beta$ is very small, different values of $C$ lead to different intervals of $\beta$ for which $r \approx 1$. For example, when $C \in [0.18,0.24]$, the final opinion assortativity $r$ is very large for a large interval of $\beta$ values; the interval of such values is smaller when $C \geq 0.3$. 


\subsection{Convergence Time} \label{sec:4.4}

\begin{figure}[t!] 
\centering
\includegraphics[width=0.9\textwidth]{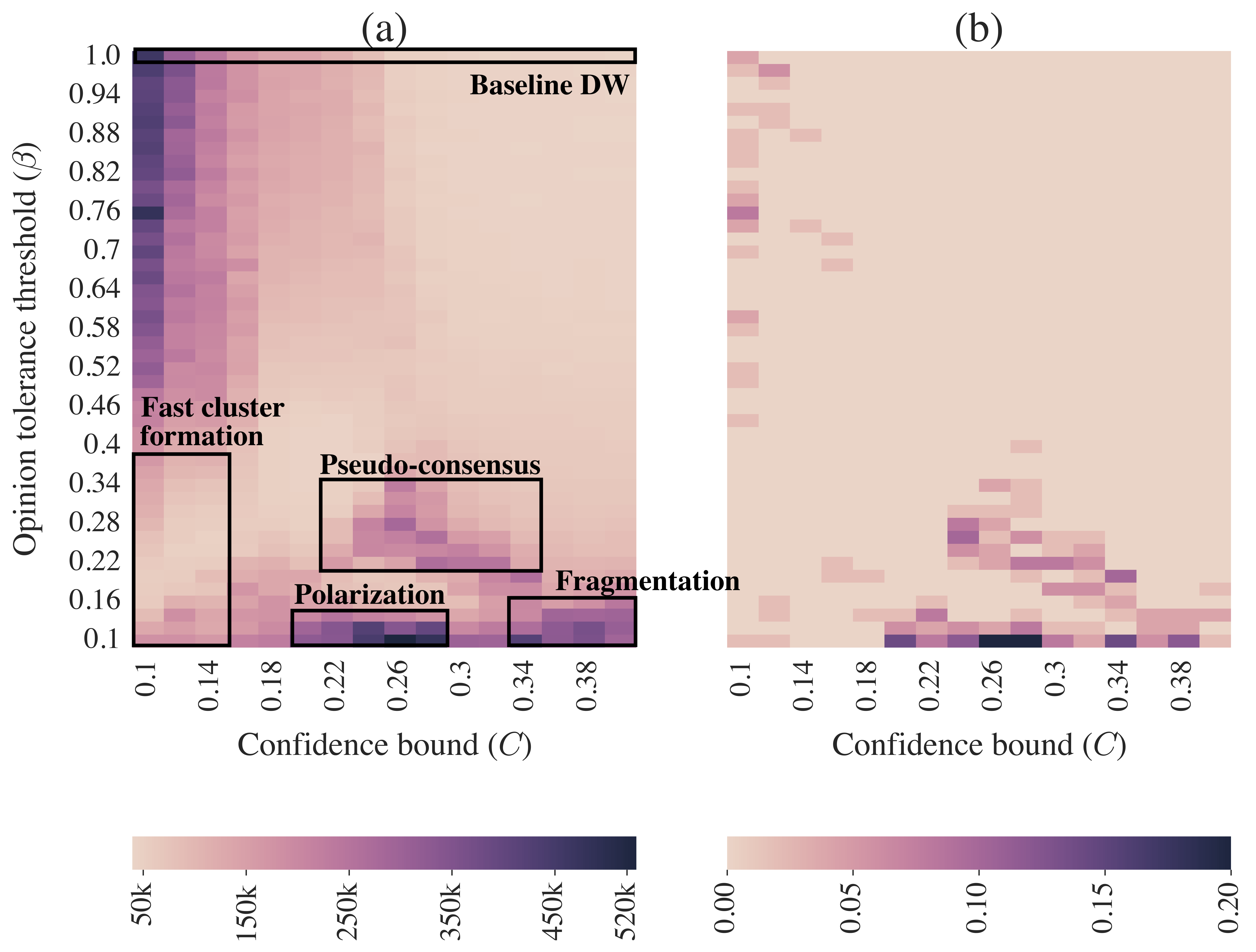}
\caption{The (a) convergence times (i.e. the numbers of time steps to reach a steady state) and (b) fractions of simulations that reach the bail-out time (i.e. they do not converge within $10^6$ steps) for our adaptive DW model. The results are means of the same set of $50$ simulations as in Figures \ref{fig:num-clusters}. An opinion tolerance threshold of $\beta = 1$ corresponds to baseline DW dynamics (i.e. without rewiring).} 
\label{fig:avg-convergence-time}
\end{figure}

We roughly organize the $(\beta, C)$ parameter plane into different steady-state regimes of our model by examining the convergence times of our simulations. In Figure \ref{fig:avg-convergence-time}(a), we show a heat map of these convergence times. It illustrates that the convergence times for intermediate values of the confidence bound $C$ are noticeably longer for small values of the opinion tolerance threshold $\beta$ than they are for the baseline case $\beta = 1$ (i.e. when there is no rewiring). We label the associated region in the heat map with the term `Polarization'. In this region, our simulations reach the bail-out time more frequently than in other regions (see Figure \ref{fig:avg-convergence-time}(b)).
 
When $C$ is small, the convergence times of our simulations are much faster when there is a lot of rewiring (i.e. for small $\beta$) than when there is not. See the bottom-left region of the heat map in Figure \ref{fig:avg-convergence-time}(a). As we discussed previously, in this regime, once the agents of a network organize into homophilic communities through rewiring, they compromise their opinions to achieve intra-community consensus. The pseudo-consensus region also has noticeably different convergence-time behavior than in the $\beta = 1$ baseline. For a fixed value of $C$, it often takes longer to converge in the former regime than in the latter regime. When agents can be influenced both by agents in their own community and by agents in other communities, their opinions fluctuate before they reach a moderate-opinion consensus through compromise. When two opinion clusters move closer to forming a pseudo-consensus, the opinions of the agents in them can keep changing by minuscule amounts before converging.


\subsection{Numerical Simulations with Larger Networks} \label{sec:4.5}

To examine the possibility that some of our adaptive DW model's key features --- such as the presence of pseudo-consensus states and the shifted phase-transition location in comparison to the baseline DW model --- arise from finite-size effects, we simulate it on networks with $N = 5000$ nodes. For these larger networks, we simulate our model for selected locations in the $(\beta, C)$ parameter plane. We again use $G(N,p)$ ER networks with $p = 0.01$, so the expected mean degree is now $\langle k \rangle = 50$. We run one set of experiments in which we fix $C \in \{0.22, 0.26, 0.28, 0.30 \}$ and vary $\beta$, and we also run a set of experiments in which we fix $\beta \in \{0.12, 0.28, 0.32, 1.0 \}$ and vary $C$. We again let $\alpha = 0.1$. We also retain the ratio $M/K = 1/5$, but we now take $M = 5$ and $K = 25$. For each examined value of $(\beta,C)$, we perform 50 simulations and calculate sample means.

In these numerical simulations, we observe several of the same key phenomena as we did in our simulations on 1000-node networks. First, we obtain the same key result for convergence times in our adaptive DW model in comparison to those in the baseline DW model (i.e. when $\beta = 1$). The convergence to steady state is faster in our adaptive DW model than in the baseline DW model when $\beta$ and $C$ are both small, and it is is slower in our adaptive model than in the baseline DW model when $\beta$ is small and $C$ is large. Second, when $\beta$ is small, the number of minor opinion clusters (which include fewer than $5$ \% of the nodes of a network) increases with $C$. Third, we again observe pseudo-consensus steady states, which again have associated long convergence times. Fourth, opinion assortativity tends to decrease with $\beta$.

Our simulations on larger networks give some insights into possible finite-size effects. We observe a transition from polarization to consensus as we increase $C$ from $C = 0.22$ to $C = 0.30$, and a phase transition appears to occur at $C'_{\beta} \approx 0.28$ for all examined values of $\beta$ (including $\beta=1$). By contrast, our simulations on $1000$-node ER networks suggested that $C_\beta'$ depends on $\beta$. (In those simulations, small values of $\beta$ shift the phase transition to values up to $C'_{\beta} \approx 0.34$.) Nevertheless, homophilic rewiring (i.e. $\beta < 1$) can still influence the process to reach consensus, which is easier to achieve in the $5000$-node ER networks in some situations but harder to achieve in these networks in other situations. On one hand, as in our simulations with 1000-node networks, homophilic rewiring can make it harder for the system to reach consensus. For example, when $C = 0.26$, almost half of our simulations with $\beta = 1$ for 5000-node networks result in consensus, but fewer than 20 \% of our simulations with $\beta = 0.1$ reach consensus. On the other hand, we also observe in our simulations on 5000-node networks that homophilic rewiring can lead to pseudo-consensus states for values of $C$ that are below the putative phase transition. For example, many simulations with intermediate values of $\beta$ result in pseudo-consensus states when $C = 0.26 < C'_{\beta} \approx 0.28$. Additionally, some simulations at the putative phase transition yield pseudo-consensus states with intermediate values of $\beta$.

There are some issues to keep in mind in our exploration of possible finite-size effects in our numerical simulations. For example, we use larger values of $M$ and $K$ for our simulations with 5000-node networks than for simulations with 1000-node networks, so we need to be cognizant that these changes can also affect our results. Although $M/K = 1/5$ in both cases, the change from $M = 1$ to $M = 5$ may speed up the rewiring process by a different amount than the change from $K = 5$ to $K = 25$ speeds up the opinion-update process. We also wonder whether or not pseudo-consensus states continue to exist in the limit $N \rightarrow \infty$.


\section{Conclusions and Discussion} \label{sec5}

We developed an adaptive bounded-confidence model (BCM) on networks that generalizes the Deffuant--Weisbuch (DW) model by allowing discordant edges to rewire based on opinion homophily. We studied our adaptive DW model on $G(N,p)$ Erd\H{o}s--R\'{e}nyi networks, and we found that it is harder for networks to achieve consensus when there is rewiring than when there is not rewiring (and hence in a baseline DW model). In one region of parameter space, we observed `pseudo-consensus' steady states with two opinion subclusters (with a minuscule difference between the opinions of the agents in the two subclusters) within a consensus opinion group. We observed that the convergence times of numerical simulations of our model tend to be long near critical values $C'_{\beta}$ of the confidence bound $C$ that separate consensus steady states from polarized and fragmented steady states. We also observed that the convergence times tend to be short when both $C$ and the opinion tolerance threshold $\beta$ are small, in contrast to the typical behavior of the standard DW model. We obtained similar numbers of major opinion clusters at steady state with and without rewiring, and we demonstrated that large values of $C$ can encourage the formation of minor opinion clusters, whose nodes tend to have extreme opinion values at the beginning of our simulations.

There are a variety of aspects of our adaptive DW model that will benefit from further explorations. For example, we focused primarily on situations in which opinion changes are faster than network rewiring (i.e. $M < K$). Our preliminary investigation (which we did not discuss previously) of situations with rewiring rates that are faster than opinion dynamics (i.e. $M > K$) reveals scenarios with very large confidence bounds (such as $C = 0.4$, which is above the phase transition when $M < K$) in which we observe polarized and fragmented steady states. That is, for large confidence bounds, it is harder for agents to achieve consensus when $M > K$ than when $M < K$. Another worthwhile direction is to further analyze the influence of moderate agents on opinion dynamics. Moreover, it is important to further study the possibility of finite-size effects, to compare different ways of selecting multiple agent pairs for possible opinion updates in a given time step, and to analyze our model in the limit of infinitely many agents. 

As with other models of opinion dynamics, our adaptive DW model includes various unrealistic assumptions, which we made for simplicity. For example, we assumed that all agents have the same confidence bound $C$ and the same opinion tolerance threshold $\beta$, and it is more realistic to incorporate heterogeneity in these parameters. Additionally, in our model, each agent that breaks a connection needs to form a new edge, but people can unfriend someone on a social network (e.g. during hotly contested political elections) without also connecting to someone else. Moreover, in our model, we did not constrain the number of times that an agent can rewire or include a core group of agents that they will never unfriend. In reality, it is possible that some people always remain connected to certain other people (e.g. family members or particularly close friends). It is worthwhile to explore relaxations of these and other assumptions of our adaptive DW model to help improve our understanding of the spread of opinions in social networks in realistic situations.


\section*{Acknowledgements}

We thank Heather Zinn Brooks and the other participants of UCLA's Networks Journal Club for helpful comments. We also thank Mark Neidengard for useful discussions and an anonymous referee for helpful comments. We acknowledge financial support from the National Science Foundation (grant number 1922952) through the Algorithms for Threat Detection (ATD) program.





\end{document}